# Monitoring of the Einstein Cross with the Nordic Optical Telescope

R. Østensen[1], S. Refsdal[2], R. Stabell[3], J. Teuber[4], P.I. Emanuelsen[5], L. Festin[5], R. Florentin-Nielsen[4], G. Gahm[6], E. Gullbring[6], F. Grundahl[7], J. Hjorth[7], M. Jablonski[5], A.O. Jaunsen[5], A.A. Kaas[5], H. Karttunen[5], J. Kotilainen[8], E. Laurikainen[8], H. Lindgren[5], P. Mähönen[9], K. Nilsson[8], G. Olofsson[6], Ø. Olsen[5], B.R. Pettersen[10], V. Piirola[9], A.N. Sørensen[5], L. Takalo[8], B. Thomsen[7], E. Valtaoja[8], M. Vestergaard[4], and T. av Vianborg[3]

[1] Institute of Mathematical and Physical Sciences, University of Tromsø, N-9037 Tromsø, Norway
[2] Hamburger Sternwarte, Gojenbergsweg 112, D-21029 Hamburg, Germany
[3] Institute of Theoretical Astrophysics, University of Oslo, Blindern, N-0315 Oslo, Norway
[4] Copenhagen University Observatory, Brorfelde, DK-4340 Tølløse, Denmark
[5] Nordic Optical Telescope, Observatorio Roque de los Muchachos, E-38700 Santa Cruz de La Palma, Spain
[6] Stockholm Observatorium, S-13336 Saltsjöbaden, Sweden
[7] Institute of Physics and Astronomy, University of Aarhus, DK-8000 Aarhus C, Denmark
[8] Tuorla Observatory, University of Turku, FIN-21500 Piikkiö, Finland
[9] Nuclear & Astrophysics Laboratory, Keble Road, Oxford OX1 3RH, U.K.
[10] Geodetic Institute, Norwegian Mapping Authority, N-3500 Hønefoss, Norway

**Abstract.** We report results from five years of monitoring of the Einstein Cross (QSO2237+0305) with the Nordic Optical Telescope. The photometry, mainly in the R and I bands, has been performed by a PSF fitting and 'cleaning' procedure, in which the four image components as well as the host galaxy and its nucleus are iteratively removed. The resulting lightcurves exhibit several microlensing features; one event may have a timescale as short as 14 days. Variations on timescales of several years are found in all four images. This becomes even more convincing when our data are combined with data published for 1986-89. No clear high amplification event was observed during the period. A brightening of all four components during 1994 is interpreted as intrinsic variation.

## 1. Introduction

After some preliminary observations in 1990, a program was started in 1991 at the Nordic Optical Telescope (NOT) on the island of La Palma for monitoring mainly

*Send offprint requests to*: R. Stabell

four well known gravitational lens systems: QSO0142-100 = UM673 (ESO GL1), QSO0957+561 (the 'classical' Double Quasar), QSO1413+117 (Clover Leaf) and QSO2237+0305 (Einstein Cross). There are many good reasons for conducting a monitoring program for gravitational lenses (Refsdal and Surdej 1992), and with its superb optics and generally good seeing conditions NOT is excellently suited for this kind of observations. The aim of the program was to obtain lightcurves for the four lens systems. The importance of well sampled and accurate lightcurves for any of these systems can hardly be overestimated. Although observations once a week was aimed at, several gaps in the data have occurred due to bad weather conditions and technical or other problems. More than forty visiting astronomers in addition to the staff at NOT have participated in the program which ended officially in 1993; many observations from 1994 and one from 1995 are, however, included in the present material.

Deferring the first three program lenses for later publications, we will here present results for the Einstein Cross, a rather unique system which has attracted observers and theoreticians alike. It consists of a barred spiral galaxy, at a redshift $z = 0.039$, in which Huchra et al. (1985) discovered a high-redshift quasar ($z = 1.695$), later shown to be quadruply lensed. Yee (1988) published accurate data for the individual quasar components based on CCD frames in three different filters obtained with the Canada-France-Hawaii Telescope. The same year Schneider et al. (1988)

presented their results based on two observations made with the Hale Telescope and three made with the Mayall Telescope. Values for the surface mass density $\kappa$ and shear parameter $\gamma$, at each of the four image positions, were determined. Similar investigations have been performed by others (Kent & Falco 1988, Webster et al. 1991). The lensing optical depth due to individual compact objects (stars) in the galaxy is found to be up to about 0.5. The light paths of the four images of the quasar pass through the bulge of the galaxy, making gravitational influence of single stars on the light rays highly probable. The Einstein Cross is therefore the perfect system for producing microlensing effects (Chang & Refsdal, 1979, Paczynski 1986, Kayser et al. 1986, Kayser & Refsdal, 1989).

Due to the high degree of symmetry of the system, the proximity of the lensing galaxy, and the small image separations, the predicted time delays between the four images are one day or less. Therefore one can more easily distinguish between intrinsic variability and microlensing effects. Several microlensing events have been observed already for this system (Irwin et al. 1989, Pettersen 1990, Corrigan et al. 1991, Racine 1992, Yee & deRobertis 1992); however, a high amplification event (HAE) remains to be detected.

The most complete lightcurve published until now is that by Corrigan et al. (1991). They presented the results from all the direct image CCD data, of sufficient quality, available at that time. Their best sampled lightcurve (R-band) includes 15 data points. The uncertainties of the data are not discussed, but in a previous paper (Irwin et al. 1989) the systematic errors caused by imperfect modeling of the foreground galaxy were estimated to be approximately 0.05 magnitudes. For a lightcurve based upon direct image deconvolution, see Houde & Racine (1994).

## 2. Observations and technical details

The present observational material was obtained with the 2.56 m Nordic Optical Telescope, at Roque de los Muchachos, La Palma, Canary Islands, Spain, from August 1990 to January 1995. During the first three years a Tektronix 512×512 CCD camera with 0.2 arcsecond pixels was the only detector available. A filter wheel immediately in front of the $LN_2$ cooled camera allowed selection of standard broad band BVRI filters plus an intermediate band birefringent filter at 550 nm. In 1994, a new 1024×1024 CCD camera with better resolution and sensitivity was installed, and most of the Einstein Cross observations after May 1994 have been made with this camera. Reasonable signal-to-noise ratios are obtained for exposure times of 4 minutes and more when the seeing is better than 1 arcsecond. Exposure times are controlled to within 0.1 seconds by a mechanical shutter. Accurate telescope tracking during exposures was ensured by auto-guiding on a nearby star. Our best images show point source profiles with FWHM = 0.5 arcsec, which clearly resolves the four optical components of QSO2237+0305 in addition to the nucleus of the foreground galaxy.

From the full data set for the Einstein Cross, frames obtained during 45 nights were of sufficient quality to be used in our investigation, resulting in 89 broad-band observations. This in effect will more than double the number of data points available and also considerably extend the time span covered. Some of the very best images have been used to construct algorithms for processing and modeling of the quasar components and the lensing galaxy. A list of the observations is presented in Table 1.

For preprocessing of our data we have applied the IRAF/ccdred package developed and maintained by NOAO (National Optical Astronomy Observatories, Tucson, Arizona). This package contains the necessary tools for performing bias correction, bad pixel and cosmic ray removal, dark current correction, and flat fielding.

Until 1993, the 512×512 CCD camera suffered from a non-linearity problem caused by a flaw in the amplifiers before A/D conversion. The non-linearity was studied by Kjeldsen (1990) who found no colour dependence nor any time variation and subsequently developed a correction algorithm which was adopted throughout. While the deviations from linearity may be as large as 10 %, the corrected values are linear to better than 1 %.

## 3. Data reductions

The Einstein Cross is a prototype of a complicated photometric system. The first prerequisite for acceptable photometry is good seeing, well below the source separations of about one arcsecond. The presence of both the point-like nucleus and the extended light distribution of the lensing galaxy adds to the well-known complications of a crowded field. Aperture photometry is of course excluded, and even PSF fitting, the method used in our work, is not trivial. Several approaches for PSF determination exist (see Janes and Heasley 1993 and references therein). We have focused upon two different methods: The semi-analytical one employed in the DAOPHOT package by Stetson (1987), and the new 'Rotate-And-Stare' empirical profile estimation described by Teuber et al. (1994). The latter paper also contains a brief account of our algorithms for PSF fitting and single source photometry. To do the overlapping source photometry, we apply the CLEAN deconvolution procedure (Teuber 1993), where the individual images are iteratively removed, together with subtraction of a de Vaucouleurs ($R^{1/4}$) galaxy model, until a satisfactory residual is obtained. The CLEAN procedure has been implemented in the IDL software package XECClean (Østensen 1994), a handy tool for interactive photometry.

A by-product of the CLEAN photometry is the set of exact positions for the individual image components. In the above procedure, the positions are obtained by applying a Maximum Likelihood centering algorithm (see

**Table 1.** List of observations

| Date | Filters | FWHM | Observer |
|---|---|---|---|
| 19.08.90 | I | 0.9" | Olsen |
| 28.08.90 | VR | 0.5" | Pettersen |
| 31.08.90 | R | 0.9" | Olsen |
| 17.09.90 | VR | 0.8" | Olsen |
| 21.09.90 | VR | 0.8" | Olsen |
| 09.11.90 | VR | 0.7" | Olsen |
| 11.11.90 | VR | 0.9" | Olsen |
| 24.05.91 | VRI | 0.7" | Emanuelsen |
| 10.07.91 | VRI | 0.9" | Kotilainen |
| 18.07.91 | RI | 0.8" | Laurikainen |
| 25.07.91 | VRI | 0.9" | Stabell |
| 17.08.91 | VRI | 0.7" | Vestergaard |
| 11.09.91 | RI | 0.8" | Piirola |
| 13.10.91 | RI | 0.7" | Hjorth |
| 16.10.91 | R | 1.1" | Pettersen |
| 30.10.91 | RI | 0.8" | Vestergaard |
| 11.06.92 | R | 1.0" | Stabell |
| 09.08.92 | RI | 1.1" | Gullbring |
| 30.08.92 | RI | 0.9" | Karttunen |
| 04.10.92 | RI | 0.9" | Thomsen |
| 12.11.92 | R | 1.0" | Olofson |
| 20.07.93 | RI | 0.7" | Valtaoja |
| 25.07.93 | RI | 0.9" | Lindgren |
| 01.08.93 | RI | 0.9" | Lindgren |
| 17.08.93 | RI | 0.8" | Lindgren |
| 23.08.93 | RI | 0.6" | Nilsson |
| 17.09.93 | RI | 0.7" | Sørensen |
| 22.09.93 | RI | 0.9" | Sørensen |
| 21.05.94 | VR | 0.6" | Jaunsen |
| 28.05.94 | VR | 0.7" | Jablonski |
| 20.06.94 | VR | 0.4" | Jablonski |
| 01.07.94 | VR | 0.5" | Jablonski |
| 03.08.94 | VRI | 0.7" | Teuber |
| 05.08.94 | VRI | 0.9" | Teuber |
| 19.08.94 | VR | 0.9" | Jablonski |
| 04.09.94 | VR | 0.7" | Takalo |
| 07.09.94 | VR | 0.8" | Takalo |
| 18.09.94 | R | 0.5" | Festin |
| 22.10.94 | R | 1.0" | Jaunsen |
| 23.10.94 | VR | 0.9" | Jaunsen |
| 24.10.94 | VR | 0.5" | Jaunsen |
| 09.11.94 | VR | 0.9" | Mähönen |
| 23.12.94 | RI | 0.6" | Jablonski |
| 30.12.94 | RI | 0.7" | Festin |
| 06.01.95 | RI | 1.0" | Festin |

Teuber et al. 1994). Table 2 gives the positions of the components determined by averaging the results from 17 observations made with the new CCD camera in 1994. For comparison, we have included the corresponding positions from Rix et al. (1992). The two sets of position are plotted in Figure 1. The small discrepancies could possibly reflect the difference in centering methods used. In Rix et al. and in previous work by other authors (Crane et al. 1991, Racine 1991, 1992, Yee 1988), no deconvolution is carried out, and the centering procedures used are not necessarily statistically optimal. Alternatively, the discrepancy might be ascribed to a slight (0.1 percent) error in the telescope/instrument scale conversion. It is important to note, however, that this unresolved *astrometric* issue does not affect the *photometric* results in any significant way.

**Table 2.** Astrometry for the three components B, C, D and galaxy nucleus N; positions are given in arcseconds relative to the A image.

| Id. | Rix et al. | | This paper | |
|---|---|---|---|---|
|  | R.A. | Dec. | R.A. | Dec. |
| B | −0.676 | 1.686 | −0.684 | 1.665 |
| C | 0.625 | 1.200 | 0.593 | 1.208 |
| D | −0.869 | 0.520 | −0.844 | 0.535 |
| N | −0.083 | 0.918 | −0.093 | 0.927 |

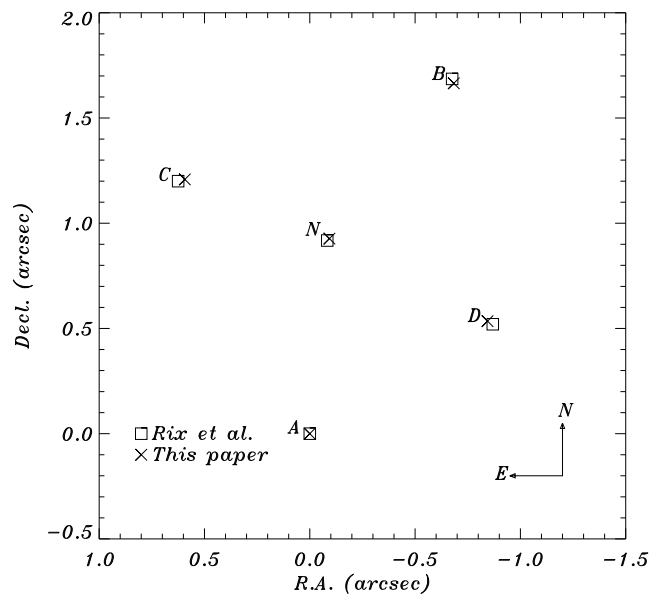

**Fig. 1.** Astrometry relative to the A component, compared to Rix et al. (1992)

## 4. Results and Discussion

Our photometric results are presented in Tables 3 - 5. The magnitudes are calibrated by means of Yee's (1988) reference star, using the transformation equations and revised values given by Corrigan et al. (1991).

In these tables, the error estimates in the rightmost column are based on both the statistical uncertainties, as derived from the XECClean fitting, and the residuals in the CLEAN'ed image. The latter contribution includes the effects from galaxy subtraction, seeing conditions, and the quality of the PSF. Zero-point errors occuring when using a calibration star in the field, with different color than the program object, can be as large as 0.02 magnitudes when changing from one detector to another. Accordingly, we have put an effective lower limit to the uncertainties of 0.03 magnitudes; the actual statistical quality is of course higher for the strong A and B components (see discussion below).

**Table 3.** V-band magnitudes

| Date (UT) | A | B | C | D | $\Delta m$ |
|---|---|---|---|---|---|
| 28.08.90 | 17.97 | 17.73 | 18.57 | 18.70 | 0.04 |
| 17.09.90 | 18.04 | 17.88 | 18.61 | 18.80 | 0.07 |
| 21.09.90 | 18.03 | 17.86 | 18.55 | 18.79 | 0.07 |
| 09.11.90 | 17.90 | 17.74 | 18.51 | 18.81 | 0.07 |
| 11.11.90 | 17.92 | 17.73 | 18.49 | 18.70 | 0.07 |
| 24.05.91 | 17.73 | 17.30 | 18.43 | 18.58 | 0.06 |
| 10.07.91 | 17.69 | 17.42 | 18.42 | 18.55 | 0.08 |
| 25.07.91 | 17.77 | 17.35 | 18.51 | 18.60 | 0.08 |
| 17.08.91 | 17.79 | 17.45 | 18.61 | 18.75 | 0.08 |
| 21.05.94 | 17.42 | 17.49 | 18.57 | 18.83 | 0.06 |
| 28.05.94 | 17.46 | 17.54 | 18.57 | 18.67 | 0.05 |
| 20.06.94 | 17.42 | 17.49 | 18.56 | 18.81 | 0.05 |
| 01.07.94 | 17.42 | 17.47 | 18.59 | 18.83 | 0.03 |
| 03.08.94 | 17.45 | 17.49 | 18.59 | 18.84 | 0.06 |
| 05.08.94 | 17.40 | 17.45 | 18.53 | 18.72 | 0.04 |
| 19.08.94 | 17.37 | 17.44 | 18.53 | 18.71 | 0.04 |
| 05.09.94 | 17.39 | 17.44 | 18.50 | 18.79 | 0.04 |
| 07.09.94 | 17.37 | 17.42 | 18.49 | 18.75 | 0.03 |
| 23.10.94 | 17.34 | 17.37 | 18.44 | 18.67 | 0.05 |
| 24.10.94 | 17.35 | 17.38 | 18.45 | 18.75 | 0.03 |
| 09.11.94 | 17.36 | 17.39 | 18.43 | 18.72 | 0.03 |

**Table 4.** R-band magnitudes

| Date (UT) | A | B | C | D | $\Delta m$ |
|---|---|---|---|---|---|
| 28.08.90 | 17.65 | 17.49 | 18.13 | 18.30 | 0.04 |
| 31.08.90 | 17.65 | 17.48 | 18.12 | 18.34 | 0.04 |
| 17.09.90 | 17.73 | 17.58 | 18.16 | 18.42 | 0.05 |
| 21.09.90 | 17.72 | 17.54 | 18.13 | 18.39 | 0.04 |
| 09.11.90 | 17.71 | 17.60 | 18.22 | 18.36 | 0.07 |
| 11.11.90 | 17.69 | 17.53 | 18.21 | 18.41 | 0.06 |
| 24.05.91 | 17.62 | 17.26 | 18.24 | 18.32 | 0.04 |
| 10.07.91 | 17.56 | 17.33 | 18.09 | 18.41 | 0.04 |
| 18.07.91 | 17.59 | 17.27 | 18.26 | 18.36 | 0.03 |
| 25.07.91 | 17.64 | 17.26 | 18.14 | 18.33 | 0.05 |
| 17.08.91 | 17.58 | 17.29 | 18.19 | 18.39 | 0.06 |
| 11.09.91 | 17.56 | 17.30 | 18.16 | 18.38 | 0.05 |
| 13.10.91 | 17.55 | 17.26 | 18.13 | 18.35 | 0.03 |
| 16.10.91 | 17.58 | 17.25 | 18.14 | 18.50 | 0.05 |
| 30.10.91 | 17.59 | 17.45 | 18.30 | 18.41 | 0.06 |
| 11.06.92 | 17.38 | 17.32 | 18.10 | 18.39 | 0.06 |
| 09.08.92 | 17.40 | 17.37 | 18.17 | 18.41 | 0.03 |
| 30.08.92 | 17.42 | 17.30 | 18.16 | 18.35 | 0.04 |
| 04.10.92 | 17.36 | 17.32 | 18.17 | 18.35 | 0.08 |
| 12.11.92 | 17.28 | 17.27 | 18.18 | 18.33 | 0.07 |
| 20.07.93 | 17.24 | 17.25 | 18.20 | 18.50 | 0.04 |
| 25.07.93 | 17.26 | 17.23 | 18.19 | 18.34 | 0.03 |
| 01.08.93 | 17.29 | 17.25 | 18.17 | 18.52 | 0.03 |
| 17.08.93 | 17.33 | 17.26 | 18.28 | 18.52 | 0.04 |
| 23.08.93 | 17.32 | 17.29 | 18.21 | 18.44 | 0.03 |
| 17.09.93 | 17.27 | 17.25 | 18.19 | 18.44 | 0.04 |
| 22.09.93 | 17.30 | 17.24 | 18.24 | 18.34 | 0.04 |
| 21.05.94 | 17.26 | 17.31 | 18.26 | 18.52 | 0.05 |
| 28.05.94 | 17.27 | 17.36 | 18.21 | 18.43 | 0.05 |
| 20.06.94 | 17.27 | 17.34 | 18.25 | 18.47 | 0.05 |
| 01.07.94 | 17.23 | 17.31 | 18.26 | 18.48 | 0.03 |
| 03.08.94 | 17.26 | 17.32 | 18.26 | 18.55 | 0.05 |
| 05.08.94 | 17.25 | 17.28 | 18.25 | 18.44 | 0.04 |
| 19.08.94 | 17.19 | 17.25 | 18.16 | 18.40 | 0.04 |
| 04.09.94 | 17.22 | 17.28 | 18.23 | 18.47 | 0.03 |
| 07.09.94 | 17.21 | 17.26 | 18.22 | 18.44 | 0.03 |
| 18.09.94 | 17.22 | 17.23 | 18.17 | 18.47 | 0.06 |
| 22.10.94 | 17.19 | 17.21 | 18.23 | 18.39 | 0.05 |
| 23.10.94 | 17.16 | 17.20 | 18.20 | 18.36 | 0.05 |
| 24.10.94 | 17.17 | 17.23 | 18.22 | 18.47 | 0.03 |
| 09.11.94 | 17.17 | 17.25 | 18.19 | 18.44 | 0.03 |
| 23.12.94 | 17.13 | 17.21 | 18.13 | 18.41 | 0.03 |
| 30.12.94 | 17.14 | 17.22 | 18.12 | 18.37 | 0.03 |
| 06.01.95 | 17.12 | 17.19 | 18.14 | 18.29 | 0.04 |

In the lightcurves (Figs. 2 and 3), we have plotted our measurements for the four components in the R and I bands, respectively. During the period in question, no *bona fide* microlensing HAE has been observed, although the somewhat uneven sampling does not exclude a possible occurrence. We find that the A image has undergone a gradual rise until marginally regaining (early 1994) its former status as the brightest component. Since we cannot see corresponding effects in the other components, this must be mainly due to microlensing effects. Note, however, that the observed increase during 1994 is seen in all four components. This must therefore be interpreted as an intrinsic variation of the source, since the theoretically predicted time delay between the components is of the order of one day. A linear regression analysis of the data from 1994 (and 06.01.95) gives a brightness increase of about 0.0006 magnitudes per day. Although this is not in itself sufficient for measuring the time delay, it is interesting to note that the correlation is rather tight, with a standard deviation for the A and B components of about 0.02 mag., which is only about half of the typical error estimates given in Tables 3-5 (average for all four components).

As regards the B component, the most interesting feature is the very significant drop of about 0.20 mag in the R-band and 0.25 in the I-band in two weeks at the end of the 1991 observational period. The magnitude is well determined before this event, and the single data point of 30th October represents three observations from the same night (two in R and one in I), all confirming the low intensity value. Unfortunately, we have no more observations from 1991; hence, it is impossible to determine the true amplitude and time-scale of this event. However, from the discussion in Refsdal & Stabell (1993), it is seen that such a sudden drop in brightness could easily be explained by microlenses with masses as small as $10^{-6}$ $M_\odot$, although this may imply an unreasonably small source size.

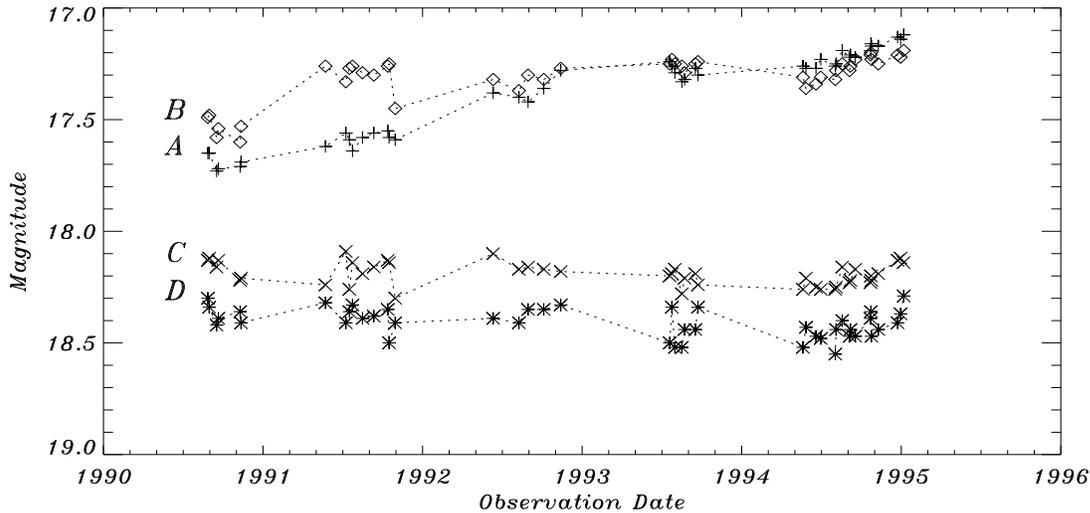

**Fig. 2.** R-band lightcurve for the four components of the Einstein Cross

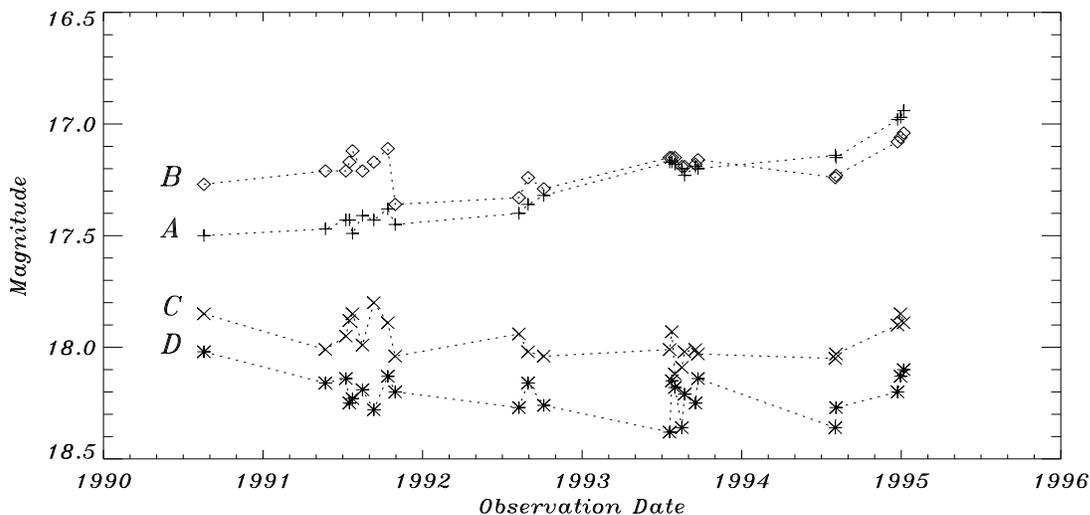

**Fig. 3.** I-band lightcurve for the four components of the Einstein Cross

In Figure 4, we have plotted our results for the four components in the R-band, together with earlier data published by Corrigan et al. (1991). One notable feature is the overall decline of the A component of roughly half a magnitude in less than a year between 1989 and 1990, as already noted by Racine (1992). An extra observation point by Racine reduces the time to about half a year. For a discussion of possible implications of this event, see Refsdal & Stabell (1993). There is a clear trend in the observations towards increased difference in brightness between the two pairs A, B and C, D, which can only be ascribed to long-term microlensing, since this difference is unaffected by intrinsic variations, considering the short time delay. Over the period 1986-1994, we note a general intensity decrease for the two weak components, approximately 0.8, 0.4 and 0.5 magnitudes for the C component, and 0.5, 0.2 and 0.3 magnitudes for the D component, in the V, R, and I bands, respectively. This dimming and rather large chromatic effects are particularly pronounced in the early data and have more or less subsided since 1990. The variations could all be due to long-term microlensing effects, but it is not possible to exclude some intrinsic variations. Microlensing effects on a time scale of ten years are to be expected, as noted by Kayser et al. (1986).

From our monitoring data, we find very nearly equal colours for components A and B ($V - R \approx 0.18, R - I \approx 0.11$). The colours for components C and D are also roughly equal ($V - R \approx 0.32, R - I \approx 0.20$). The data from 1994 which were obtained with the improved CCD camera, show an almost constant colour for both pairs. If colour effects from microlensing are negligible, the reddening is in agreement with the approximate $\lambda^{-1}$ extinction law given by Nadeau et al. (1991) and leads to an extinction difference in the V band of about 0.6 magnitudes between the two pairs, in good agreement with the value found by Fitte and Adam (1994). For the R and I bands, the extinction differences are around 0.45 and 0.35 mag., respectively.

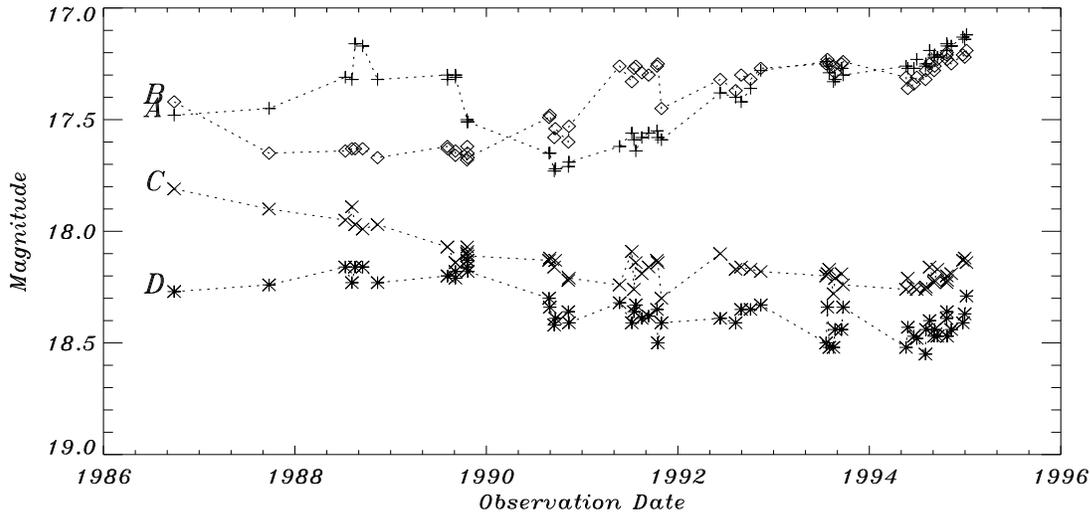

**Fig. 4.** Complete R-band lightcurve for the period 1986-1995 based on data from this paper and from Corrigan et al. (1991).

**Table 5.** I-band magnitudes

| Date (UT) | A | B | C | D | $\Delta m$ |
|---|---|---|---|---|---|
| 19.08.90 | 17.50 | 17.27 | 17.85 | 18.02 | 0.05 |
| 24.05.91 | 17.47 | 17.21 | 18.01 | 18.16 | 0.06 |
| 10.07.91 | 17.43 | 17.21 | 17.95 | 18.14 | 0.04 |
| 18.07.91 | 17.43 | 17.17 | 17.88 | 18.25 | 0.04 |
| 25.07.91 | 17.49 | 17.12 | 17.85 | 18.23 | 0.05 |
| 17.08.91 | 17.41 | 17.21 | 17.99 | 18.19 | 0.04 |
| 11.09.91 | 17.43 | 17.17 | 17.80 | 18.28 | 0.05 |
| 14.10.91 | 17.38 | 17.11 | 17.89 | 18.13 | 0.04 |
| 30.10.91 | 17.45 | 17.36 | 18.04 | 18.20 | 0.06 |
| 09.08.92 | 17.40 | 17.33 | 17.94 | 18.27 | 0.04 |
| 30.08.92 | 17.36 | 17.24 | 18.02 | 18.16 | 0.04 |
| 04.10.92 | 17.32 | 17.29 | 18.04 | 18.26 | 0.08 |
| 20.07.93 | 17.17 | 17.15 | 18.01 | 18.38 | 0.04 |
| 25.07.93 | 17.17 | 17.15 | 17.93 | 18.15 | 0.03 |
| 01.08.93 | 17.18 | 17.15 | 18.12 | 18.18 | 0.03 |
| 17.08.93 | 17.20 | 17.19 | 18.09 | 18.36 | 0.05 |
| 23.08.93 | 17.23 | 17.19 | 18.02 | 18.21 | 0.03 |
| 17.09.93 | 17.19 | 17.18 | 18.01 | 18.25 | 0.04 |
| 22.09.93 | 17.20 | 17.16 | 18.03 | 18.14 | 0.05 |
| 03.08.94 | 17.14 | 17.24 | 18.05 | 18.36 | 0.05 |
| 05.08.94 | 17.15 | 17.23 | 18.03 | 18.27 | 0.05 |
| 23.12.94 | 16.98 | 17.08 | 17.90 | 18.20 | 0.03 |
| 30.12.94 | 16.97 | 17.06 | 17.85 | 18.13 | 0.03 |
| 06.01.95 | 16.94 | 17.04 | 17.89 | 18.10 | 0.04 |

The low sampling rate and the uncertainty of the measurements does not exclude a possible flickering in the lightcurves of the order of 0.05 magnitudes or less. Such flickering is expected to occur if the microlens effect is due to low mass objects (Refsdal & Stabell 1993). Better sampling of the lightcurves, with several observations per night to improve accuracy, is essential for resolving such features. Frequent monitoring of the Einstein Cross, with well resolved images, yielding accurate lightcurves, is in all circumstances invaluable for the determination of quasar characteristics as well as extinction properties and the detailed mass distribution of the compact objects in the bulge region of the lensing galaxy.

*Acknowledgements.* We wish to extend our thanks to all visiting astronomers at NOT who have participated in the monitoring program. We also would like to thank the referee C. Vanderriest for very valuable comments and suggestions. Two of us (RØ and JT) want to thank Institute of Theoretical Astrophysics, University of Oslo, for hospitality and support during the preparation of this work.

This research was also supported by the Norwegian Research Council.